# BEYOND GEOMETRICAL OPTICS AND BOHMIAN PHYSICS: A NEW EXACT AND DETERMINISTIC HAMILTONIAN APPROACH TO WAVE-LIKE FEATURES IN CLASSICAL AND QUANTUM PHYSICS

A. Orefice, R. Giovanelli and D. Ditto

Universita' di Milano - DI.PRO.VE. - Via Celoria, 2 - 20133 - Milano (ITALY)

**Abstract**  The indeterministic character of physical laws is generally considered to be the most important consequence of quantum physics. A deterministic point of view, however, together with the possibility of well defined Hamiltonian trajectories, emerges as the most natural one from the analysis of the time-independent Helmholtz-like equations encountered both in Classical Electromagnetism and in Wave Mechanics. In the case of particle beams a suitable pattern of trajectories is provided (for any set of boundary conditions) by a set of dynamical laws containing the classical ones as a simple limiting case.

## 1- Introduction

Due to the disdainful attitude of influent Founding Fathers such as Heisenberg and Einstein, the main *alternative* interpretation of Quantum Mechanics - the "hidden variables" point of view proposed by de Broglie **[1-3]** and Bohm **[4,5]** - did not enter in the mainstream of Physics, and was forced to develop into a separate, somewhat esoteric and almost heretical  "church" **[6]**.

In the present (simple, but not necessarily simplistic) work an approach bearing some analogies with that "heretical" standpoint, but with a quite different general philosophy, arrives at a set of deterministic equations describing the quantum motion of a particle beam and containing the classical dynamical laws as a particular case, thus suggesting that the standard probabilistic treatment of quantum features may constitute the best approach when a detailed information is lacking or unnecessary, but does not reflect an intrinsically indetermined nature of physical reality.

As is well known (and as we shall see in the next Section), the **Helmholtz** equation, describing a wide family of monochromatic wave-like phenomena, may be reduced to a system of *two coupled equations* (eqs. (5)  of the present paper).

The *first* of these equations is usually truncated, by neglecting the term coupling it to the *second* equation. In such an *incomplete form* it provides, by itself alone, the set of "rays" which characterizes the so-called "geometrical optics approximation". No further contribution to the ray geometry is given, in this limit, by the *second* of eqs.(5).

In the present paper the coupled equation system (5) is shown **for the first time** to lead, *without any omission or approximation*, to an **exact** Hamiltonian ray-tracing set of equations (eqs.(25) of the present paper), providing the **exact** description of a family of classical wave-like phenomena much wider than that allowed by the geometrical optics approximation and including, for instance, wave diffraction and interference.

**For the first time in the history of Physics** the term dropped from the *first* of eqs.(5) in the usual context of geometrical optics is shown to be of crucial importance for the determination of the structure and motion of the wave beam, and an equal importance is shown to be attached to the *second* of eqs.(5).

**For the first time** it is shown that a correlation exists between the rays of a beam, due to a term $(\frac{\partial}{\partial \underline{x}} \frac{\nabla^2 R}{R})$ acting *perpendicularly* to the ray velocities, and determining therefore both their geometry and motion.

**For the first time in the history of Physics** the Hamiltonian system (25) holding for *classical* ray is shown to *coincide exactly* with the *quantum*-dynamical Hamiltonian system (eqs.(20) of the present paper) deduced from the *time-independent* **Schrödinger** equation (which is itself a **Helmholtz**-like equation), leading to a unique Hamiltonian system of the general form (26).

The term neglected in the context of geometrical optics is shown to coincide, in its turn, with the *quantum potential* described by Bohm and de Broglie:

**for the first time**, therefore, in the 50 years elapsed from Bohm's works, the basic character of the "force" deduced from this *quantum potential* (i.e. the fundamental property of being *transversal with respect to to the particle trajectories*) is discovered, stressed and exploited.

**For the first time** such a deterministic Hamiltonian system, in the form (28), is numerically solved for an arbitrary beam consisting, indifferently, **either** of *classical* rays **or** of *quantum* particle trajectories, and its diffractive behaviour is clearly shown by this solution.

The basic intuition allowing this **harvest of novel results** stems from the formal coincidence between the **time independent Schrödinger** equation (14) and the **Helmholtz** equation (1): any other approch would lead, to say the least, to useless complication and confusion. Both Holland's book **[6a]**, indeed,  and his UK.ARXIV paper **[6b]** are affected by such a complication and confusion, fully explaining by itself the failure of Bohm's approach to be accepted and developed in 50 years.

**2 - The Helmholtz equation**

In order to establish the mathematical formalism to be extended, later on, to the **quantum** treatment of a particle beam, let us start from a **classical** case of wave-like behaviour.

Although many kinds of waves would lend themselves to the considerations we have in mind here, we shall refer, in order to fix ideas, to a monochromatic *electromagnetic* wave beam, with a time dependence $\div exp\ (i\omega t),$ travelling through an isotropic and inhomogeneous dielectric medium. Its basic features are accounted for by the Helmholtz equation

$$\nabla^2 \psi + (n\ k_0)^2 \psi = 0\ , \qquad (1)$$

where $\psi$ represents any component of the electric or magnetic field, $n(x,y,z)$ is the refractive index of the medium and

$$k_0 \equiv \frac{2\pi}{\lambda_0} = \frac{\omega}{c}\ , \qquad (2)$$

with obvious meaning of $\lambda_0$ and $c$. The phase velocity is given, in its turn, by

$$v_{ph}(x,y,z) = c / n(x,y,z). \qquad (3)$$

Because of its time-independence, eq.(1) doesn't directly describe, of course, any propagation phenomenon: it only determines, together with the boundary conditions, the *fixed* space frame where propagation occurs.

By performing the quite general replacement

$$\psi(x,y,z) = R(x,y,z) e^{i\ \varphi(x,y,z)}\ , \qquad (4)$$

with real $R(x,y,z)$ and $\varphi(x,y,z)$, and separating the real from the imaginary part, eq.(1) splits into the well known **[7]** and strictly equivalent system of coupled equations

$$\begin{cases} (\underline{\nabla}\varphi)^2 - (nk_0)^2 = \dfrac{\nabla^2 R}{R} \\ \underline{\nabla} \cdot (R^2\ \underline{\nabla}\varphi) = 0 \end{cases} \qquad (5)$$

where $\underline{\nabla} \equiv \partial/\partial\underline{x} \equiv (\partial/\partial x, \partial/\partial y, \partial/\partial z),\ \nabla^2 = \dfrac{\partial^2}{\partial x^2} + \dfrac{\partial^2}{\partial y^2} + \dfrac{\partial^2}{\partial z^2}$ and the

second equation expresses the constance of the flux of the vector $R^2 \underline{\nabla}\varphi$ along any tube formed by the lines of $\underline{\nabla}\varphi$ itself, i.e. normally to the surfaces $\varphi(x,y,z) = const.$ When the space variation length, $L$, of the amplitude $R(x,y,z)$ may be assumed to satisfy the condition $k_0 L >> 1$, the first of eqs.(5) is well approximated by the *eikonal equation*

$$(\underline{\nabla}\varphi)^2 \cong (nk_0)^2\ , \qquad (6)$$

decoupled from the second of eqs.(5), and allowing the so-called "geometrical optics approximation", which describes wave propagation in terms of "rays" travelling along the field lines of $\underline{\nabla}\varphi$, directly provided by eq.(6).

Let us finally define the wave vector

$$\underline{k} = \underline{\nabla} \varphi \ , \qquad (7)$$

and conclude the present Section by recalling Fermat's variational principle, according to which any optical ray travelling between two points *A,B* shall follow a trajectory satisfying the condition

$$\delta \int_A^B k \, ds = 0 \ , \qquad (8)$$

where $k = |\underline{k}|$ and *ds* is an element of a (virtual) line connecting *A* and *B*.

### 3 - The time-independent Schrödinger equation

The classical motion of a mono-energetic beam of non-interacting particles of mass *m* through a force field deriving from a potential energy $V(\underline{x}) \equiv V(x,y,z)$ not explicitly depending on time may be described for each particle of the beam, as is well known, by means of the so-called "reduced" (or "time-independent") Hamilton-Jacobi equation **[7]**

$$(\underline{\nabla} S)^2 = 2\,m\,(E - V) \ , \qquad (9)$$

where *E* is the total energy, and one of the main properties of the function *S(x,y,z)* is that the particle momentum is given by

$$\underline{p} = \underline{\nabla} S \ . \qquad (10)$$

Recalling Maupertuis' variational principle

$$\delta \int_A^B p \, ds \equiv 0 \ , \qquad (11)$$

with $p = |\underline{p}|$, the formal analogy between eqs.(6-8) on one side, and eqs.(9-11) on the other, suggests, as is well known, that the classical particle trajectories constitute the geometrical optics approximation of an equation analogous to eq.(1), which is immediately obtained by means of the substitutions

$$\begin{cases} \varphi = \dfrac{S}{a} \quad \text{and therefore} \quad \underline{k} = \underline{\nabla}\varphi = \dfrac{\underline{\nabla} S}{a} = \dfrac{\underline{p}}{a} \\[6pt] k_0 \equiv \dfrac{2\pi}{\lambda_0} = \dfrac{\sqrt{2mE}}{a} \equiv \dfrac{p_0}{a} \\[6pt] n^2(x,y,z) = 1 - \dfrac{V(x,y,z)}{E} \end{cases} \qquad (12)$$

where the parameter *a* represents a constant *action* whose value is *a priori* arbitrary - as far as the relations (12) are concerned - but is imposed by the history itself of Quantum Mechanics :

$$a = \hbar \cong 1.0546 \times 10^{-27} erg \cdot s. \qquad (13)$$

The equation obtained from the Helmholtz equation (1) by means of the substitutions (12) and (13) takes up the form

$$\nabla^2 \psi + \frac{2m}{\hbar^2}(E-V)\psi = 0, \qquad (14)$$

which is the standard time-independent Schrödinger equation. By applying now the same procedure leading from eq.(1) to eqs.(5), and assuming therefore

$$\psi(x,y,z) = R(x,y,z)e^{\,i\,S(x,y,z)/\hbar} \qquad (15)$$

eq.(14) splits into the strictly equivalent **[8]** coupled system

$$\begin{cases} (\underline{\nabla} S)^2 - 2m(E-V) = \hbar^2 \dfrac{\nabla^2 R}{R} \\ \underline{\nabla} \cdot (R^2 \, \underline{\nabla} S) = 0 \end{cases} \qquad (16)$$

By taking the gradient of the first of eqs.(16) we get moreover

$$(\frac{\underline{\nabla} S}{m} \cdot \underline{\nabla}) \frac{\underline{\nabla} S}{m} + \frac{\underline{\nabla} V}{m} = \frac{\hbar^2}{2\,m^2} \underline{\nabla}(\frac{\nabla^2 R}{R}). \qquad (17)$$

Eq.(17), together with the second of eqs.(16), is usually interpreted as describing, in the "classical limit" $\hbar \to 0$ (whatever such a limit may mean), a "fluid" of particles with mass *m* and velocity $\dfrac{\underline{\nabla} S}{m}$, moving in an external potential *V(x,y,z):* an interpretation consistent with the probabilistic character ascribed to the Schrödinger equation.

**4 - Hamiltonian approach**

Let us now observe that, by simply maintaining eq.(10), the first of eqs.(16) may be written in the form of a generalized, time-independent Hamiltonian

$$H(\underline{p}, \underline{x}) \equiv \frac{p^2}{2m} + V - \frac{\hbar^2}{2m}\frac{\nabla^2 R}{R} = E, \qquad (18)$$

including a new and crucial term $\dfrac{\hbar^2}{2m}\dfrac{\nabla^2 R}{R}$, to be commented later on.

By differentiating eq.( 18):

$$\frac{\partial H}{\partial \underline{x}} \cdot d\,\underline{x} + \frac{\partial H}{\partial \underline{p}} \cdot d\,\underline{p} = 0 \qquad (19)$$

it is seen to be associated with a Hamiltonian system of dynamical equations of the form

$$\begin{cases} \dfrac{d\underline{x}}{dt} = \dfrac{\partial H}{\partial \underline{p}} = \dfrac{\underline{p}}{m} \\ \dfrac{d\underline{p}}{dt} = -\dfrac{\partial H}{\partial \underline{x}} = -\dfrac{\partial}{\partial \underline{x}}[V(\underline{x}) - \dfrac{\hbar^2}{2m}\dfrac{\nabla^2 R}{R}] \end{cases} \quad (20)$$

We must recall, of course, the presence of the second of eqs. (16), which may be written in the form

$$\underline{\nabla} \cdot (R^2 \underline{\nabla} S) \equiv 2R\underline{\nabla}R \cdot \underline{\nabla}S + R^2 \underline{\nabla} \cdot \underline{\nabla}S = 0, \quad (21)$$

which we shall reduce to the stronger condition

$$\begin{cases} \underline{\nabla} \cdot \underline{\nabla} S = 0 \\ \underline{\nabla} R \cdot \underline{\nabla} S = 0 \end{cases}. \quad (22)$$

Clearly enough, whenever the first of eqs.(22) may be assumed to hold (an assumption, indeed, which does not appear to be too much restrictive), the second one is authomatically entailed. The values of the function *R(x)* are therefore constant (i.e. "transported") along the field lines of $\underline{p} \equiv \underline{\nabla} S$, to which $\underline{\nabla} R$ turns out to be perpendicular. A basic consequence of this property is the fact that the gradient $\dfrac{\partial}{\partial \underline{x}} \dfrac{\nabla^2 R}{R}$ computed along each trajectory remains perpendicular to the trajectory itself (i.e. tangent to the wave-front), without acting on the **amplitude** of the particle velocity (but modifying, in general, its **direction**). The only possible amplitude changes could be due to the presence of an external potential *V(x)*.

Thanks to its constance along each trajectory, moreover, the function *R(x)*, once assigned on the launching surface from where the particle beam is assumed to start, may be numerically built up step by step, together with its derivatives, in the whole region spanned by the motion of the beam.

Let us point out that a completely analogous set of equations may be obtained in the electromagnetic case of Sect.2. From the first of eqs.(5) we obtain in fact, multiplying it, for convenience, by the factor $\dfrac{c}{2k_0}$, and recalling eq.(7), the relation

$$D(\underline{r}, \underline{k}) \equiv \dfrac{c}{2k_0}[k^2 - (nk_0)^2 - \dfrac{\nabla^2 R}{R}] = 0, \quad (23)$$

whose differentiation

$$\dfrac{\partial D}{\partial \underline{x}} \cdot d\underline{x} + \dfrac{\partial D}{\partial \underline{k}} \cdot d\underline{k} = 0 \quad (24)$$

suggests the ray-tracing system

$$\begin{cases} \dfrac{d\underline{x}}{dt} = \dfrac{\partial D}{\partial \underline{k}} = \dfrac{c\,\underline{k}}{k_0} \\ \dfrac{d\underline{k}}{dt} = -\dfrac{\partial D}{\partial \underline{x}} = \dfrac{c}{2k_0}\dfrac{\partial}{\partial \underline{x}}[(nk_0)^2 + \dfrac{\nabla^2 R}{R})] \end{cases} \quad (25)$$

where a ray velocity $\underline{v}_{ray} = \dfrac{c\,\underline{k}}{k_0}$ is implicitly defined. We may observe that $v_{ray} = c$ for $k = k_0$, and $v_{ray}\,v_{ph} \cong c^2$ in the geometrical optics limit.

Because of the transverse character, mentioned above, of the gradient $\dfrac{\partial}{\partial \underline{x}}\dfrac{\nabla^2 R}{R}$, the amplitude $v_{ray}$ of the ray velocity **in vacuo** remains equal to $c$ all along the trajectory, since such a transverse term may only modify the **direction**, but not the **amplitude**, of the wave vector $\underline{k}$.

The system (25) presents the same general properties of the system (20), and completely avoids the approximation of geometrical optics, although fully retaining the idea of electromagnetic "rays" travelling along the field lines of $\underline{k} \equiv \underline{\nabla}\varphi$.

Before proceeding to the discussion and treatment of the dynamical system (20), a quite expedient step is the passage to the new, dimensionless variables $\underline{\xi}$, $\underline{\rho}$, $\tau$ defined as the ratio between $\underline{x}$, $\underline{p}$ and $t$, respectively, and

$\lambda_0 \equiv 2\pi\hbar/p_0$ for the space variables,

$p_0 \equiv (2mE)^{1/2}$ for the momentum variables (so that $\rho_0 = 1$), and

$\dfrac{\lambda_0}{p_0/m}$ for the time variable.

The equation system (20) takes up therefore the form

$$\begin{cases} \dfrac{d\underline{\xi}}{dt} = \underline{\rho} \\ \dfrac{d\underline{\rho}}{dt} = -\dfrac{\partial}{\partial \underline{\xi}}[\dfrac{V(\underline{\xi})}{2E} - \dfrac{1}{8\pi^2}\,G(\underline{\xi})] \end{cases} \quad (26)$$

with

$$G(\underline{\xi}) = \dfrac{1}{R}(\dfrac{\partial^2 R}{\partial \xi^2} + \dfrac{\partial^2 R}{\partial \eta^2} + \dfrac{\partial^2 R}{\partial \zeta^2}); \quad \underline{\xi} \equiv (\xi,\eta,\zeta) \equiv (x/\lambda_0,\,y/\lambda_0,\,z/\lambda_0). \quad (27)$$

It may be observed that no direct reference is present, in the dimensionless form (26) assumed by the dynamical system (20), to the mass of the moving particles, and not even to $\hbar$. The same dimensionless form would be taken up, indeed, by the ray tracing system (25) - relevant to the electromagnetic case - by simply assuming $\tau = \dfrac{c\,t}{\lambda_0}$ and replacing $\underline{\rho}$ with $\underline{k}/k_0 \equiv \underline{v}_{ray}/c$, and $V(x,y,z)/E$ with $[1 - n^2(x,y,z)]$, in agreement with the relations (12).

Once assigned on the launching surface of the beam, the function $G(\underline{\xi})$ may be numerically determined step by step, in principle, together with its derivatives, by means of an interpolation process iterated along the full set of trajectories of the beam and connecting each step to the previous ones.

This function, due to the wave amplitude distribution of the beam on the advancing wave-front, turns out to be the same - in correspondence with the same boundary conditions - for electromagnetic rays as well as for material particles, although it has obviously nothing to do, in the electromagnetic case, with quantum features. In its absence, however, the system (26) would simply describe the "classical" motion of each particle of the beam. Due to the small coefficient $\dfrac{1}{8\pi^2}$, the transverse gradient $\dfrac{\partial G}{\partial \underline{\xi}}$ acts along the trajectory pattern in a soft and cumulative way: a fact granting the main justification for omitting such a term, as is done both in classical dynamics and in the geometrical optics approximation.

The trajectory pattern, in its turn, is a stationary structure determined at the very outset in a way somewhat reminding the spirit of classical variational principles, such as the ones of Fermat and Maupertuis. For any set of boundary conditions imposed to the function $R(\underline{x})$ on the launching surface of the beam, and for any assigned refractive medium (or force field), the system (26) provides both a "weft" of "rails" and a motion law to which particles (or rays) are deterministically bound, showing no trace of probabilistic features. The basic point to be stressed is the influence of the full set of boundary conditions on the motion of each particle (or ray) of the beam: a point which concerns, however, Wave Mechanics as well as Classical Electromagnetism (together with whatever phenomenon may be described in terms of Helmholtz-like equations). Any attempt, indeed, to apply the time independent Schrödinger equation to a single particle (not belonging to a beam) would not appear to be more plausible than the application of the Helmholtz equation (1) to a single ray.

The modern point of view of Quantum Mechanics on indeterminism has nothing to do, as is well known, with the naive idea of a disturbance due to the observer, which would imply an underlying deterministic situation "blurred" by the observation device. Indeterminism is currently conceived, in fact, as an intrinsic natural property, forbidding, even in principle, to assign a definite trajectory to a moving particle, and reserving to the observer the subtle role of inducing (in general) the collapse of the observed system from a superposition of its possible states into a single one of them, according to well defined probabilities.

Contrary to this point of view, however, each particle (or ray) of the beam turns out to be conceivable, on the basis of the present analysis, as *starting and remaining* on a well definite trajectory. Such a trajectory belongs to a pattern which is *a priori* fixed, as a whole, by the properties of the crossed medium and by the values assigned to the beam amplitude distribution R(x,y,z) on the launching surface.

The system (20), in conclusion, provides a set of dynamical laws which replace - and contain as a limiting case, when the transverse gradient $\frac{\partial}{\partial \underline{x}} \frac{\nabla^2 R}{R}$ may be assumed to be negligible - the classical ones. Let us observe that the possibility of neglecting such a term, and of obtaining therefore a classical-looking description, may turn out to be limited to a simple portion (typically, the central part) of a beam. In striking divergence from the classical dynamical laws, however, the new set of equations, because of its equivalence with a Helmholtz-like equation, requires in general the full set of boundary conditions for the determination of each trajectory of the beam.

## 5 - Wave-like features in Hamiltonian form

Although an accurate and general numerical treatment lies beyond the aims of the present paper, we propose here the application of the equation system (26) to the propagation of a collimated beam injected at $\zeta = 0$ parallel to the $\zeta$-axis, and centered at $\xi = 0$, in order to simulate wave diffraction through a single slit.

The problem may be faced by taking into account for simplicity sake (but with no substantial loss of generality) either a *particle beam* in the absence of external fields (***V = 0***), or an *electromagnetic beam* in vacuum (***n$^2$ = 1***), with a geometry allowing to limit the computation to the trajectories lying on the $(\xi,\zeta)$-plane.

The Hamiltonian system (26) takes up therefore the form

$$\begin{cases} \dfrac{d\xi}{d\tau} = \rho_x \\ \dfrac{d\zeta}{d\tau} = \rho_z \\ \dfrac{d\rho_x}{d\tau} = \dfrac{1}{8\pi^2} \dfrac{\partial}{\partial \xi} G(\xi,\zeta) \\ \dfrac{d\rho_z}{d\tau} = \dfrac{1}{8\pi^2} \dfrac{\partial}{\partial \zeta} G(\xi,\zeta) \end{cases} \qquad (28)$$

with

$$G(\xi,\zeta) = \frac{1}{R}\left(\frac{\partial^2 R}{\partial \xi^2} + \frac{\partial^2 R}{\partial \zeta^2}\right); \quad \rho_x(\zeta=0)=0; \quad \rho_z(\zeta=0)=\rho_0=1, \tag{29}$$

and a suitable amplitude distribution $R(\xi,\zeta=0)$ (from whose normalization the function G is obviously independent) imposed at $\zeta = 0$.

Because of the transverse nature of the gradient of $G(\xi,\zeta)$, the *amplitude* of the vector $\rho$ remains unchanged (in the absence of external fields and/or refractive effects) along each trajectory, leading therefore to the relation

$$\rho_z = \sqrt{\rho_o^2 - \rho_x^2} \equiv \sqrt{1-\rho_x^2}, \tag{30}$$

which may advantageously replace the fourth equation of the Hamiltonian system (28). Two possible models of the amplitude distribution $R(\xi,\zeta=0)$ are obtained by assuming

- a Gaussian distribution centered at $\xi = 0$, in the form

$$R_1(\xi;\zeta=0) \div e^{-(\frac{x}{w_0})^2} \equiv e^{-\varepsilon^2 \xi^2} \tag{31}$$

(with constant $w_0$ and $\varepsilon = \frac{\lambda_0}{w_0} \leq 1$), a functional form suggested by its smooth analytical behaviour; or

- an algebraic distribution, in the form

$$R_2(\xi;\zeta=0) \div \frac{1}{1+(\frac{x}{w_0})^{2N}} \equiv \frac{1}{1+(\varepsilon\xi)^{2N}} \tag{32}$$

(with integer $N$), which allows to represent even a quite flat central region, widening with increasing $N$. We show in **Fig.1** both the distributions $R_1$ and $R_2$, with $\varepsilon = 0.1$ and $N = 1$, and in **Fig.2** the corresponding functions

$$G_{1,2}(\xi;\zeta=0) = \frac{1}{R_{1,2}} \frac{d^2 R_{1,2}}{d\xi^2} \tag{33}$$

determining the launching conditions at $\zeta=0$. It is seen that rather similar distributions $R_{1,2}$ may lead to quite different $G_{1,2}$ and therefore to quite different trajectory patterns. In our preliminary computations the functions $G_{1,2}(\xi;\zeta>0)$ are built up step by step by means of a 3-points Lagrange interpolation. As predicted by the standard diffraction theory **[9]**, no "fringe" is found in the Gaussian case of **Fig.3** (due to the fact that the Fourier transform of the distribution $R_1$ consists of another Gaussian function), while "fringes" appear (in the form of gathering trajectories) in **Fig.4** for the algebraic initial distribution $R_2$, focusing closer to the launching plane for higher vaues of $\varepsilon$. We shall not discuss here the specific form of these fringes, since the basic result to be pointed out is their very appearance in the context of our Hamiltonian approach.

No further difficulty would be encountered in the case of *two* beams, injected parallel to the $\zeta$-axis at $\zeta = 0$ and centered, on the $\xi$-axis, at two symmetrical points $\xi = \pm \xi_0$, in order to simulate both their diffraction and their interference through a double slit.

**6 – Discussion and conclusions**

A certain analogy may be observed between the results of the present work and the ones previously published by one of the Authors (A.O.) in a quite different context **[10-12]**. Another obvious analogy is found with Refs.**[13,14]**, based on Bohm's approach, which are hindered, however, by the absence of a clear formulation and of a suitable analysis of the relevant Hamiltonian motion laws.

Our opinion, in fact, is that Bohm did not convince the scientific community because he did not stress, or perhaps did not notice, the implications, holding even beyond the quantum case, of the *time-independent* (and therefore Helmholtz-like) Schrödinger equation, allowing a much more direct, clear and powerful insight than the *time-dependent* one. While, in particular, the term $\dfrac{\hbar^2}{2m}\dfrac{\nabla^2 R}{R}$ of eq.(24) has the dimensions and the behaviour of a potential field exerting a real (transverse) force on the particle beam, the corresponding term in eq.(29), concerning an electromagnetic ray beam, has a quite different nature, but leads to a strictly similar "weft" of trajectories. A proper analysis of the deterministic Hamiltonian system in its general form (32) reveals in fact that the deviations of a particle (or wave) beam from classical dynamics (or geometrical optics) are entirely due, in any case, to the role of the gradient $\dfrac{\partial G}{\partial \underline{\xi}}$, tangent to the advancing wave-front of the beam and due to its amplitude distribution on such a surface.

We may conclude the present work by suggesting that, contrary to a well established opinion, a **statistical description** of the behaviour of a particle (or wave) beam, although representing a convenient approximation when a fully detailed information is lacking or unnecessary, doesn't supply, in principle, the most exact possible approach.

We may also state, however, that (although deserving further investigation) the Hamiltonian set of equations providing a **deterministic description** of such a behaviour is not more surprising than any other set of motion laws, and could allow an unusual insight into quantum reality.

**FIGURE CAPTIONS**

**Fig.1** - Plot of the amplitude distributions $R_{1,2}$ assigned to the beam on the launching plane $\zeta = 0$, for $\varepsilon = \dfrac{\lambda_0}{w_0} = 0.1$,

  in the Gaussian case of eq.(31) (continuous line)
  in the algebraic case of eq.(32), with $N=1$ (dotted line)

**Fig.2** - Plot of the initial functions $G_{1,2}$ of eq. (33) corresponding to the distributions $R_{1,2}$ of FIG.1

**Fig.3** - Trajectory pattern on the $(\xi,\zeta)$-plane, in the Gaussian case of FIG.1(a). The beam is truncated at $\zeta = 700$, in order to limit it to its most interesting part.

**Fig.4** - Trajectory pattern on the $(\xi,\zeta)$-plane, in the algebraic case of FIG.1(b)

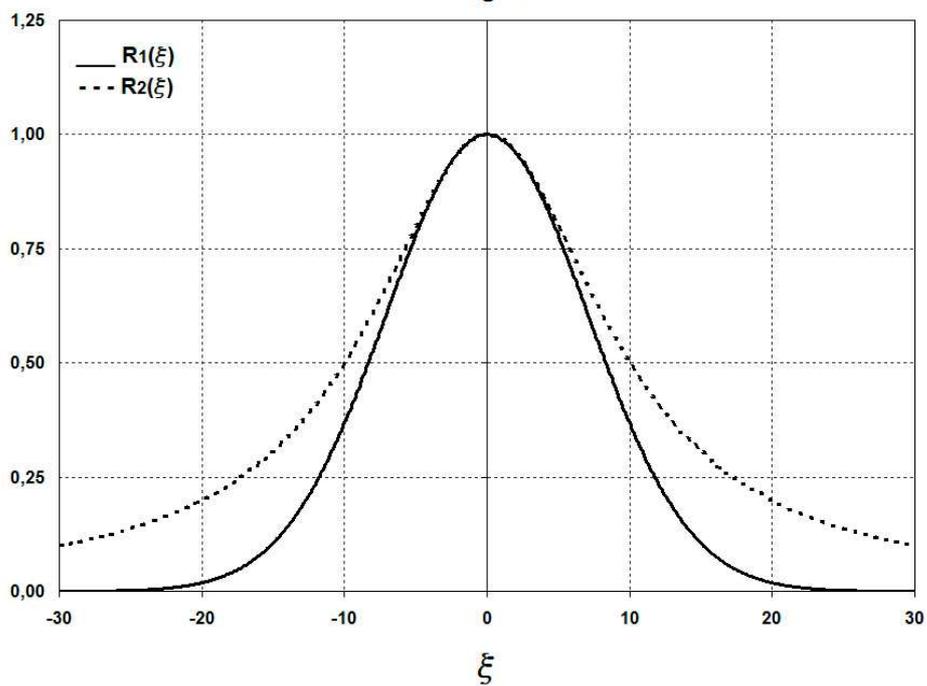

Fig. 1

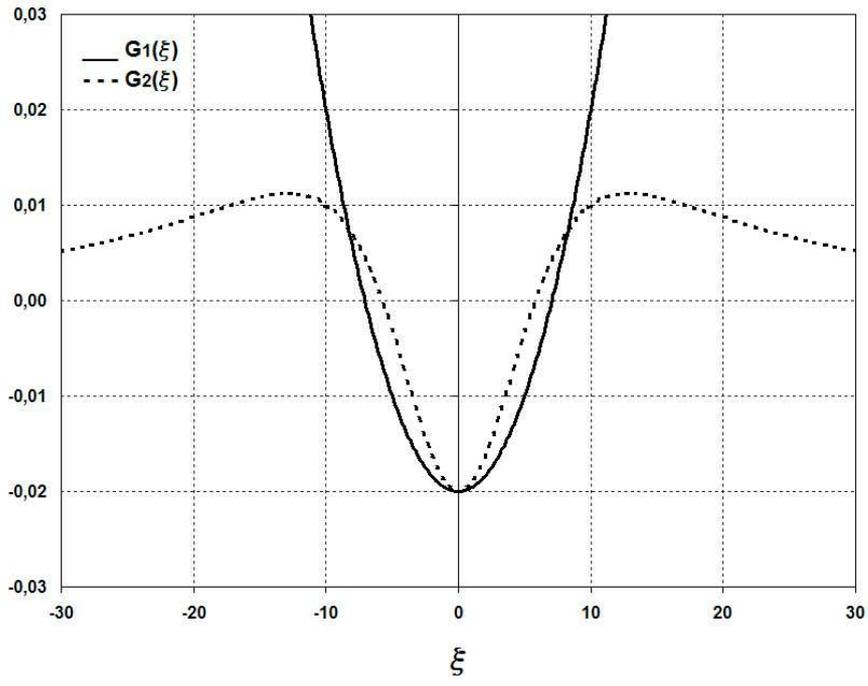

**Fig. 2**

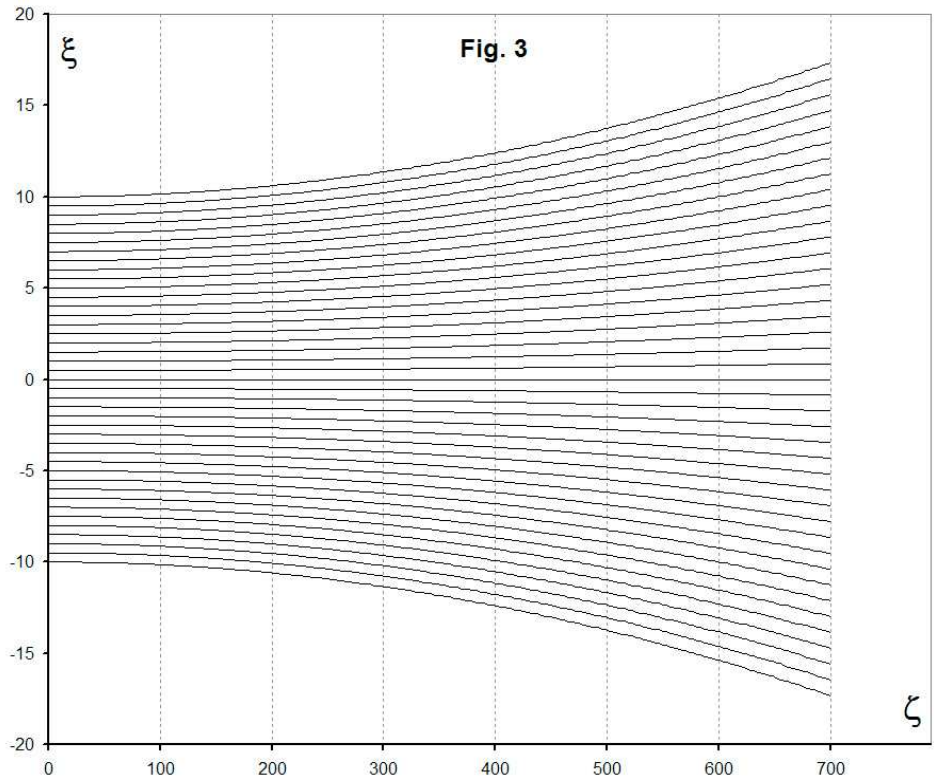

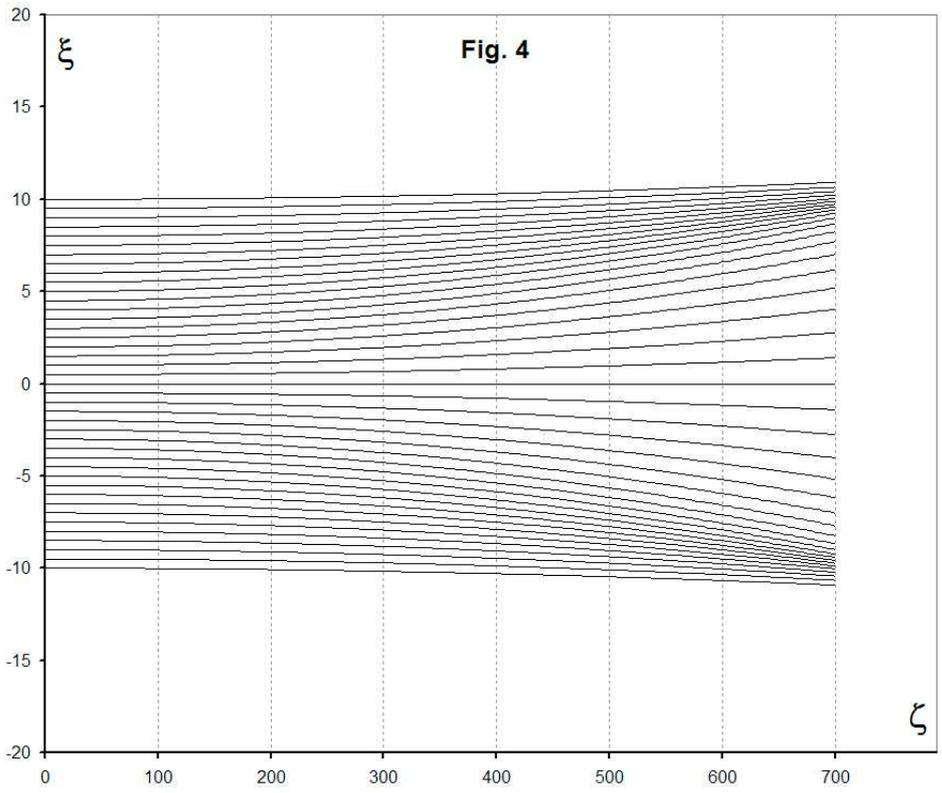

Fig. 4